\documentclass[conference]{IEEEtran}

\usepackage{cite}
\usepackage{amsmath,amssymb,amsfonts}
\usepackage{algorithmic}
\usepackage{graphicx}
\usepackage{textcomp}
\usepackage{xcolor}
\usepackage{multirow}
\usepackage{multicol}

\newtheorem{thm}{\bf Theorem}

\title{Secure Decentralized Pliable Index Coding}

\author{%
\IEEEauthorblockN{%
Tang Liu and Daniela Tuninetti\\%
University of Illinois at Chicago, Chicago, IL 60607 USA, \\
Email: {\tt tliu44, danielat@uic.edu}\\%
}%
}

\begin{document}

\maketitle

\begin{abstract}
This paper studies a variant of the Pliable Index CODing (PICOD) problem, i.e., an index coding problem where a user can be satisfied by decoding {\it any} message that is not in its side information set, where communication is {\it decentralized}, i.e., it occurs among users rather than by the central server, and {\it secure}, i.e., each user is allowed to decode only one message outside its side information set and
must not be able to collect any information about any other message that is not its decoded one.
Given the difficulty of the general version of this problem, this paper focuses on the case where the side information sets are `$s$~circular shifts', namely, user $u$'s side information set is the set of messages indexed by $\{u, u+1, \ldots, u+s-1\}$ for some fixed $s$ and where the indices are intended modulo the cardinality of the message set. 
This particular setting 
has been studied in the `decentralized non-secure' and in the `centralized secure' settings, thus allows one to quantify the cost of decentralized communication under security constraints on the number of transmissions.
Interestingly, the decentralized vs the centralized secure setting incurs a multiplicative gap of approximately~three. 
This is in contrast to the cases without security constraint, where the multiplicative gap is known to be at most~two. 
\end{abstract}

\section{Introduction}

Index Coding (IC) consists of one transmitter with $m$ independent messages and $n$ users. 
The users are connected to the transmitter through an error-free broadcast link. 
The users have side information sets locally available to them, which are subsets of the $m$ messages.
Each user has a pre-determined message as its desired message to decode. 
The transmitter has knowledge of all users' side information sets and desired messages, which it uses to generate the codewords that enable every user to decode its desired message.
Users decode based on the received codeword from the transmitter and the messages in their side information set.
The question is to find the minimum number of transmissions/code-length such that every user can decode its desired messages successfully. 
In this paper we study the {\it decentralized, secure and pliable} IC problem, which is motivated by three variants of IC: 
Pliable Index CODing (PICOD), 
decentralized IC, and 
secure IC. 

\paragraph{Pliability}
The PICOD problem is motivated by scenarios where the desired messages for the users are not pre-determined. 
Such situations include Internet radio, streaming services, online advertisement system, etc. 
In these cases, the transmitter can leverage the freedom of choosing the desired messages for the users, together with the side information sets at the users, to minimize the cost of transmission.
PICOD was proposed in~\cite{BrahmaFragouli-IT1115-7254174}, 
where the system includes a single transmitter, $m$ message, and $n$ users with message side information sets.
Different from IC, in the PICOD each user is satisfied whenever it can decode a message that is not in its side information set. 
The goal in the PICOD is to find the assignment of desired messages for the users and the corresponding transmission strategy that lead to the shortest possible code-length.

From the achievability side, results from~\cite{BrahmaFragouli-IT1115-7254174,polytime_alg_picod,constant_frac_satisfactory} show that PICOD affords an exponential code-length reduction compared to IC under the linear encoding constraint.
From the converse side under linear constraint, the work in~\cite{polytime_alg_picod} provides a lower bound on the required number of transmissions in order for the randomly generated PICOD problem to be satisfied. In~\cite{picod_it_converse}, we derived the information theoretical optimal code-length for some PICOD's with symmetric structure in the side information sets (among which the class `PICOD with circular-arc side information' of interest in this paper) by leveraging novel combinatorial arguments. 
In~\cite{picod_absent_u1,picod_absent_u2} the Authors also derived information theoretical converse bounds by explicitly leveraging  the `absent users' in the system.

\paragraph{Decentralized Communication}
The decentralized IC is motivated by 
 peer-to-peer and ad-hoc network where there is no central transmitter/server.
Here, the codewords are generated by the users based their side information set and sent through a time-sharing noiseless broadcast channel.
The goal in decentralized IC is to find the shortest code-length that allows all users to decode their desired message.

The decentralized IC is a special case of the multi-sender IC~\cite{the_single_uniprior_index_coding_problem} and of the distributed IC (where  the system has $m$ messages and  $2^m-1$ servers with different message sets available for encoding)~\cite{capacity_thm_for_distributed_ic}. 
Recently, the decentralized IC has been studied in~\cite{embedded_ic} under the name \emph{embedded} IC (where the information available for encoding at the servers is the same as the side information sets at the users), where it is shown that
the optimal number of transmissions under linear encoding constraint does not increase by more than a multiplicative factor of~2 when turning a linear code for the IC to a linear code for embedded IC.
A task-based solution (i.e., a one-short scheme) for the embedded IC was discussed in~\cite{tb_sol_embedded_ic}.

We studied the decentralized PICOD in~\cite{decentralized_picod}, where we proved the information theoretical optimal number of transmissions for all those centralized PICOD's we had solved in~\cite{picod_it_converse}; interestingly, in these cases we showed that the multiplicative gap between the optimal number of transmissions of the centralized vs the decentralized setting is often very close to or exactly equal to~1, meaning that decentralized transmission imposes a very minimal cost in terms of network load.

\paragraph{Secure Communication}
Security in IC has been studied from several perspectives. 
IC with an eavesdropper, who has a limited access to the side information sets and to the transmitted codeword, was proposed in~\cite{security_ic_with_sideinfo_eavesdropper};
in such a model the transmitter needs to satisfy all users while preventing the eavesdropper from getting any new information from the transmissions. 
Another studied model is `security against other users' where the transmitter wants to prevent the users from knowing the content of the desired messages of the other users from the received broadcast codewords. In~\cite{private(secure)_ic} the cases of (i) strong security (a user does not learn any information from the set of the non-desired messages) with secure key and (ii) weak security (a user may learn some information from the set of the non-desired messages but does not infer any information on the non-desired messages, also referred to as `individual security') without secure key were investigated.

Recently, the Authors of~\cite{on_picod} studied the case of `weak security against other users' in PICOD's with `$s$~circular shift' side information set structure, namely, user $u$'s side information set is the set of messages indexed by $\{u, u+1, \ldots, u+s-1\}$ for some fixed $s$ and where the indices are intended modulo the cardinality of the message set $m$. 
In~\cite{individual_sec_picod} we generalized the problem setup of~\cite{on_picod} and provided linear codes whose length is at most one more than a converse bound under the constraint of linear encoding. 
Our results for centralized PICOD with `$s$~circular shift' side information structure demonstrate a multiplicative gap between secure and non-secure versions of the problem that is not bounded in general; in particular, for $s\geq m/2$ the two cases have the same information theoretic optimal code-length; however, for $1\leq s <m/2$ and under the constraint of linear encoding, the minimal number of transmissions for the secure PICOD is lowered bounded by 
essentially $m/(2s)$, while the non-secure PICOD can always be satisfied by at most 2 transmissions.

{\bf Contribution and Paper Organization:}
The \emph{decentralized secure PICOD} studied in this paper is the `individual secure' version of the decentralized PICOD, motivated by the communication system without central transmitter such as peer-to-peer networks. In practice, security is a very important factor in peer-to-peer networks, since such networks usually consist of many anonymous users, and malicious users can get into the networks comparatively easily.  Security against other users in the systems guarantees that the files that are shared over the network will not fall into the hands of malicious users. 


Given the difficulty of the PICOD in general, we focus here on the case of `$s$~circular shift' side information set structure as in~\cite{on_picod,individual_sec_picod}.  
We show that, under a linear encoding constraint, several cases that are feasible in the centralized setting become infeasible in decentralized setting.
Our bounds reveal: 
\begin{enumerate}

\item
On the one hand, when $\frac{m}{m-s}\in \mathbb{Z}$, i.e., one transmission can satisfy all users in the centralized non-secure PICOD setting, the information theoretical optimal number of transmissions is $\frac{m}{s}$, 
which coincides with the optimality result for the decentralized case without security. 
Thus, in this decentralized setting security comes for free.

\item
On the other hand, when $\frac{m}{m-s} \not\in \mathbb{Z}$, i.e., two transmission can satisfy all users in the centralized non-secure PICOD setting, things are very different. 
Under linear encoding constraint, we show the converse bound $\ell^{\star}\geq \frac{3m}{2s}$, which seems to indicate that we need roughly 3 times more transmission compared to the case of centralized secure PICOD with linear encoding. 
This converse bound is shown to be tight in some cases. 
The multiplicative gap between the centralized and decentralized secure PICODs is strictly larger than the one between centralized and decentralized non-secure PICODs with the same side information set structure.
This shows a fundamental difference when we impose security constraints in decentralized settings.

\item
Our achievability result does not cover all decentralized secure PICOD's with `$s$~circular shift' side information sets, namely,
at the time of submitting this paper we had not find a general scheme for all odd values of message size $m$.
We observe that for odd $m$ and $\frac{m}{m-s}\notin \mathbb{Z}$, there are many infeasible cases  under the linear encoding constraint. 
However, not all odd $m$ and $\frac{m}{m-s}\notin \mathbb{Z}$ are infeasible; for example, the case $(m,s)=(11,8)$ is feasible.
The feasibility for general odd $m$ is subject of current investigation.
 
\end{enumerate}

The rest of the paper is organized as follows. 
Section~\ref{sec:model} introduces the system model;
Section~\ref{sec:main_result_and_discuss} provides the main results and discussion;
Section~\ref{sec:infeasible} proves the infeasibility result of Theorem~\ref{thm:infeaibility};
Section~\ref{sec:converse} and Section~\ref{sec:achievability} prove the converse and achievability parts for Theorem~\ref{thm:gap_3}, respectively.

\begin{table*}
\centering
\caption{Summary of results for PICOD's with circular shift side information sets.}
\begin{tabular}{ |c|c|c|c| } 
\hline
\multirow{1}{*}{}    & Without security & With security \\
\hline
 \multirow{3}{*}{}   &&
 Infeasible cases: $\frac{m}{m-s}\notin \mathbb{Z}$, odd $m$, $s=1$ or $s=m-2$.
 \\
 Centralized & $\ell^{\star}_{\mathrm{it}}=\begin{cases}
1, & \frac{m}{m-s} \in \mathbb{Z},\\
2, & \frac{m}{m-s} \notin \mathbb{Z}.
 \end{cases}$ 
 & 
 $\ell^{\star}_{\mathrm{it}}=\begin{cases}
1, & \frac{m}{m-s} \in \mathbb{Z},\\
2, & \frac{m}{m-s} \notin \mathbb{Z} \text{ and } s > m/2.
 \end{cases}$ 
 \\
 && 
$ \lceil \lfloor\frac{m}{s} \rfloor /2 \rceil 
\leq  
\ell^{\star}=\begin{cases}
\lceil \lfloor\frac{m}{s} \rfloor /2 \rceil , & \frac{m}{m-s}\notin \mathbb{Z}, s<\frac{m}{2}, \frac{m}{s} \in \mathbb{Z},\\
\lceil \lfloor\frac{m}{s} \rfloor /2 \rceil +1, & \frac{m}{m-s}\notin \mathbb{Z}, s<\frac{m}{2}, \frac{m}{s} \notin \mathbb{Z}.
 \end{cases}$ 
 \\
\hline
\multirow{3}{*}{}   &&
Infeasible cases: $\frac{m}{m-s}\notin \mathbb{Z}$, $s=1,2$ or $s=3,m-2$ with odd $m$
\\
Decentralized& 
 $\ell^{\star}_{\mathrm{it}}=\begin{cases}
\frac{m}{s}, & \frac{m}{m-s} \in \mathbb{Z},\\
2, & \frac{m}{m-s} \notin \mathbb{Z}.
 \end{cases}$ 
& 
$\ell^{\star}_{\mathrm{it}}=m/s, \frac{m}{m-s} \in \mathbb{Z}$\\ 
&&
$ \frac{3m}{2s}
\leq  
\ell^{\star}=\begin{cases}
\frac{3m}{2s}, & \frac{m}{m-s}\notin \mathbb{Z}, \frac{m}{2s} \in \mathbb{Z},\\
\frac{m}{2}+2-\lceil \frac{s}{2} \rceil, & \frac{m}{m-s}\notin \mathbb{Z}, \text{even } m.
 \end{cases}$ 
\\
\hline
\end{tabular}
\label{table:result_summary}
\end{table*}

\section{System Model}
\label{sec:model}

Throughout the paper we use the following notation. 
For integers $1 \leq a_1\leq a_2 $ we let $[a_1:a_2] := \{a_1,a_1+1,\ldots,a_2\}$, and $[a_2]:=[1:a_2]$.
A capital letter as a subscript denotes set of elements whose indices are in the set, i.e., $W_A:=\{w_a : w_a\in W, a\in A\}$.
For two sets $A$ and $B$, $A\setminus B$ is the set that consists all the elements that are in $A$ but not in $B$.

In this paper we study the decentralized secure PICOD problem with $m$ messages and with `$s$~circular shift' side information sets defines as follows.
\begin{enumerate}

\item 
$m\in\mathbb{N}$ users and no central transmitter. The user set is denoted as $U := \left\{ u_{1},u_{2},\ldots,u_{m} \right\}$.

\item 
$m$ messages.  
The messages are of $\kappa \in \mathbb{N}$  independent and uniformly distributed bits. 
The message set is denoted as $\mathcal{W} := \left\{ w_{1},w_{2},\ldots,w_{m} \right\}$.

\item 
User $u_i, i\in[m],$ knows the messages indexed by its side information set $A_i:= [i:(i+s-1)\mod{m}]$. 
The collection of all side information sets, denoted as $\mathcal{A} := \{A_{1},A_{2},\ldots,A_{n}\}$, is assumed globally known at all users.
For valid setup we have $s \in [m-1]$, i.e., the users have some but not all messages as side information.

\item 
A shared noiseless broadcast channel connects all users in the system.  
The users broadcast their codewords to all the other users one at a time.

\item 
The codewords are generated by each user based on their side information set.
In other words, the overall transmission is $x^{\ell \kappa} :=\{x^{\ell_1 \kappa}, \dots, x^{\ell_n \kappa}\}$ where codeword $x^{\ell_j \kappa}$ is generated by the $j$-th user as 
\begin{align}
      x^{\kappa\ell_j} := \mathsf{ENC}_{j} (W_{A_j}, \mathcal{A}), \ \forall j\in [m],
\end{align}
for some function  $\mathsf{ENC}_{j}$. 

The total code-length is $\ell := \sum_{j\in[m]} \ell_j$ and

\item 
The decoding function at the $j$-th user is 
\begin{align}
    \widehat{w}_{j}
    := \mathsf{DEC}_j(W_{A_j},x^{\ell \kappa}), \ \forall j\in [m],
\end{align}
for some function  $\mathsf{DEC}_{j}$. 

A decoding success is declared for user $u_j$ if $\widehat{w}_{j}=w_{d_j}$ for some $d_j\in [m]\setminus A_j$.
That is, user $u_j$ decodes a message that is outside its own side information set.

\item 
The decoding at the $j$-th user must also be `individually secure' meaning it must satisfy  
\begin{align}
	\label{eq:secure_constraint}
     I( W_i  | x^{ \kappa\ell}, W_{A_j}, \mathcal{A} )=0, \
     \forall i\in [m]\setminus{(\{d_j\} \cup A_j)}.
\end{align}

\item
Given $(m,s)$, we aim to find the smallest $\ell$ such that the decoding is successful and individually secure at all users. 
We assume $\kappa$ can be arbitrarily large for asymptotic analysis. 
We indicate the optimal code-length as $\ell^{\star}_{\mathrm{it}}$ (subscript `it' sands for `information theoretically optimal'). 
If we restrict the encoding functions to be linear maps, the optimal code-length is denoted by $\ell^{\star}$  (without any subscript, so as not to clutter the notation).
\end{enumerate}

\section{Main results and Discussion}
\label{sec:main_result_and_discuss}

Our main results in this paper are as follows.

\begin{thm}[\bf Infeasible cases]
	\label{thm:infeaibility}
	For the decentralized secure PICOD with circular shift side information sets and linear encoding, in the following cases
	it is not possible to satisfy all users while maintaining the security constraint:
	\begin{enumerate}
    \item $s=1, m\geq 3$;
    \item $s=2, m\geq 5$;
    \item $s=3$ and odd $m$;
    \item $s=m-2$ and odd $m$.
   \end{enumerate}
\end{thm}

\begin{thm}[\bf Converse bound with multiplicative gap of 3 compared to the centralized setting]
\label{thm:gap_3}
    For the decentralized secure PICOD with circular shift side information sets,
    when $ m/(m-s) \in \mathbb{Z} $ we have $\ell^{\star}_{\mathrm{it}}=m/s$,   
    otherwise
    \begin{align}
    \frac{3m}{2s} 
    \leq 
    \ell^{\star} 
    \leq  
    \begin{cases}
    \frac{3m}{2s}, &  m/2s \in \mathbb{Z} ,\\
    \frac{m}{2}+2-\lceil\frac{s}{2}\rceil, &   m/2s \notin \mathbb{Z} \text{ and even } m.
    \end{cases}    
    \end{align}
    
\end{thm}

Table~\ref{table:result_summary} summarizes known results for the PICOD problem with circular shift side information sets in four setups (centralized vs. decentralized transmission, and with vs. without security constraint).
We can see that, for the case $m/(m-s) \in \mathbb{Z}$, the minimal number of transmissions is the same as the case without security constraint. 
In this case the security constraint does not make any difference. 
However, the case $m/(m-s) \not\in \mathbb{Z}$ behaves differently:
\begin{itemize}
   
    \item 
    In decentralized PICOD, many feasible non-secure cases become infeasible because of the security constraint.
    Compared to the centralized setting, we have more infeasible cases in the decentralized setting. 
    For instance, the centralized case with $s=2$ is feasible but becomes infeasible in the decentralized setting.
    
    \item 
    Our general converse bound under linear encoding constraint is $\ell^\star \geq 3m/(2s)$.
    The multiplicative gap to the centralized secure PICOD when $s< m/2$ is roughly~3. 
    Note that the gap between the centralized~\cite{picod_it_converse} and decentralized~\cite{decentralized_picod} PICOD without security constraint 
    is at most~2.
    Our results show that when security constraints are imposed, the cost of decentralization for PICOD becomes substantially larger.

    \item 
    We can also quantify the impact of decentralized communication on non-secure vs. secure PICOD.
    In the centralized setting, security constraints change the linear optimality result only when $ s < m/3$.
    In the decentralized setting, security constraints change the linear optimality for at least for $s < 3m/4$, which is strictly larger than the above. 
    
    \item We remark that our results on decentralized secure PICOD is incomplete at the time of submitting this paper.     
     Our general achievable scheme for even $m$  does not exactly match the converse bound in general.      
     For odd $m$, there are many cases for which we can neither prove or disapprove feasibility at this point;
     for example, for $m=11$ the cases $s=1,2,3,9$ are infeasible while $s=10,8$ are feasible, but the other cases are still open.
     Nonetheless, our results already show a fundamental difference between decentralized and centralized secure PICOD. 
     Completing the characterization of the decentralized secure PICOD is part of ongoing work.
\end{itemize}


\section{Infeasible cases}
\label{sec:infeasible}
In this section we 
show that the cases in Theorem~\ref{thm:infeaibility} do not have a feasible linear solutions.

We note that in the proofs below we do not limit ourselves to scalar or one-shot schemes. 
In general, the transmissions can be linear combinations of fractions of messages, since each message consists $\kappa$ bits. 
Therefore, one transmission may not allow users to decode a complete message but just a part of it. 
However, by the security constraint, a user can not decode any parts for the messages that are not its desired message. 
If a user decodes a part of one message, the message is the desired message of the user. 
Then by the converse argument, the user will eventually decode its desired message and be able to mimic the other user. 

\paragraph{Case $s=1$, $m\geq 3$} 
In this case each user only has one message in its side information set. 
Since in the decentralized setting the codewords are generated based on the local message knowledge,  
we can assume without loss of generality that user $u_1$ does the first transmission.
For the first transmission, $u_1$ can only generate a codeword based on its side information $w_1$, and thus without loss of generality it transmits $w_1$.
This will allow all other users to decode $w_1$.
Once $w_1$ has been decoded, all other users who have $w_1$ in their side information set can mimic user $u_1$ and thus
decode message $w_{d_1}$ (where $d_1$ is the index of the desired message by user $u_1$).
When $m\neq 3$, there are more than two users who can mimic user $u_1$ and there is at least one user who can decode more than one messages. 
This case is thus infeasible.

Note that this argument is true even without the linear encoding constraint. Therefore the result here is  information theoretical.

\paragraph{$s=2$, $m\geq 5$}
\label{par:infeasible_s=2}
With linear encoding, the codewords are linear combinations of the locally available messages.
The first transmission is thus a linear combination of at most two messages.
We observe that a linear combination of two consecutive messages violates the security constraint. 
This is so because of the following. Without loss of generality assume the first transmission is done by user $u_2$. 
On the one hand, user $u_2$ can send a linear combination of messages $w_2$ and $w_3$;
by receiving this transmission, user $u_1$ (who has $w_2$) can decode $w_3$ and user $u_3$ (who has $w_3$) can decode $w_2$;
now both users $u_1$ and $u_3$ 
can now mimic user $u_2$ and decode $w_{d_2}$; 
since $ A_1\cap A_3= \emptyset$, $w_{d_2}$ can not be in the side information sets of both user $u_1$ and user $u_3$,
therefore, $u_1$ or $u_3$ must be able to decode one more message outside their side information set which violates the security constraint.
On the other hand, sending an uncoded message one at a time is not secure by \cite[Proposition 1]{individual_sec_picod} not even in the centralized case. 
We conclude that there is no feasible linear solution in this case.

\paragraph{$s=3$ and odd $m$}
\label{par:infeasible_s=3}
The transmission can only be a linear combination of two messages with adjacent indices. 
One transmission determine the desired messages of two users. 
The same message can not be involved in two different transmissions. 
Therefore  the involved messages of all transmissions are disjoint and linear combinations of the transmissions are not useful in terms of decoding.
$\ell$ transmissions then always satisfy  $2\ell$ users. 
However, the number of users $m$ is odd in this case. 
There will be always at least one unsatisfied user. 
Otherwise the security constraint will be violated.
We conclude there is no feasible solution for this case.

Note that for either $s=2, m\geq 5$ or $s=3,m\text{ odd}$, the argument holds true for any invertible mapping, thus not necessarily a linear code. 
We have however not been able to derive yet a fully information theoretic converse (i.e., no restriction on the encoding map).


\paragraph{$s=m-2$ and odd $m$}


This case is infeasible since it is infeasible in the centralized case without constraint of linear encoding.
Thus it is also infeasible in the decentralized case. 

\section{Converse bound}
\label{sec:converse}

In this section we prove the converse bound $\ell^{\star}\geq \frac{3m}{2s}$ under the linear encoding constraint for  $\frac{m}{m-s}\notin \mathbb{Z}$.
We construct a `chain of desired message pairs' which provides an inequality.
We then derive a lower bound based on the maximum number of satisfied users. 
By combining these two inequalities we have the desired converse bound.

\paragraph{Desired Message Pairs}
With linear encoding, each transmission is a 
linear function of the messages in the side information set of the transmitting user (i.e., we neglect the messages that a user may have already decoded thanks to previous transmissions by other users). 
This is without loss of generality because if the transmission involves codewords sent in previous time slots, we can do the transmission by just sending the part that is only involving the messages in the side information set and the rest of the users will add the contribution of the previous transmissions themselves.

Let us consider the codeword generator matrix $G$ of the overall linear code. 
The $i$-th transmission is the $i$-th row of $G$, and is denoted as $g_i$.
Let $\mathsf{Span}(G)$ denote the row vector linear span of $G$.
Under the decentralized setting, each transmission involves the messages that are in one user's side information set.
By \cite[Proposition 1]{individual_sec_picod}, $\mathsf{Span}(G)$ does not include any standard basis vector. 
That is, each transmission involves at least two messages.
Let $b_i\leq s-1$ be the range of the messages that are involved in $i$-th transmission. 
Explicitly, the $i$-th row of $G$ can have at most $b_i$ nonzero elements, in the range $[a_i:a_i+b_i-1\mod{m}]$, between the $a_i$-th and $(a_i+b_i-1)\mod{m}$-th elements.
By the argument in Section~\ref{par:infeasible_s=2},  the transmission can not involve the first and the last messages in the side information set at the same time.
Therefore, the messages that are involved in one transmission are in the range $b_i \in[2:s-1]$.

\begin{figure}
  \centering
  \includegraphics[width=0.5\columnwidth]{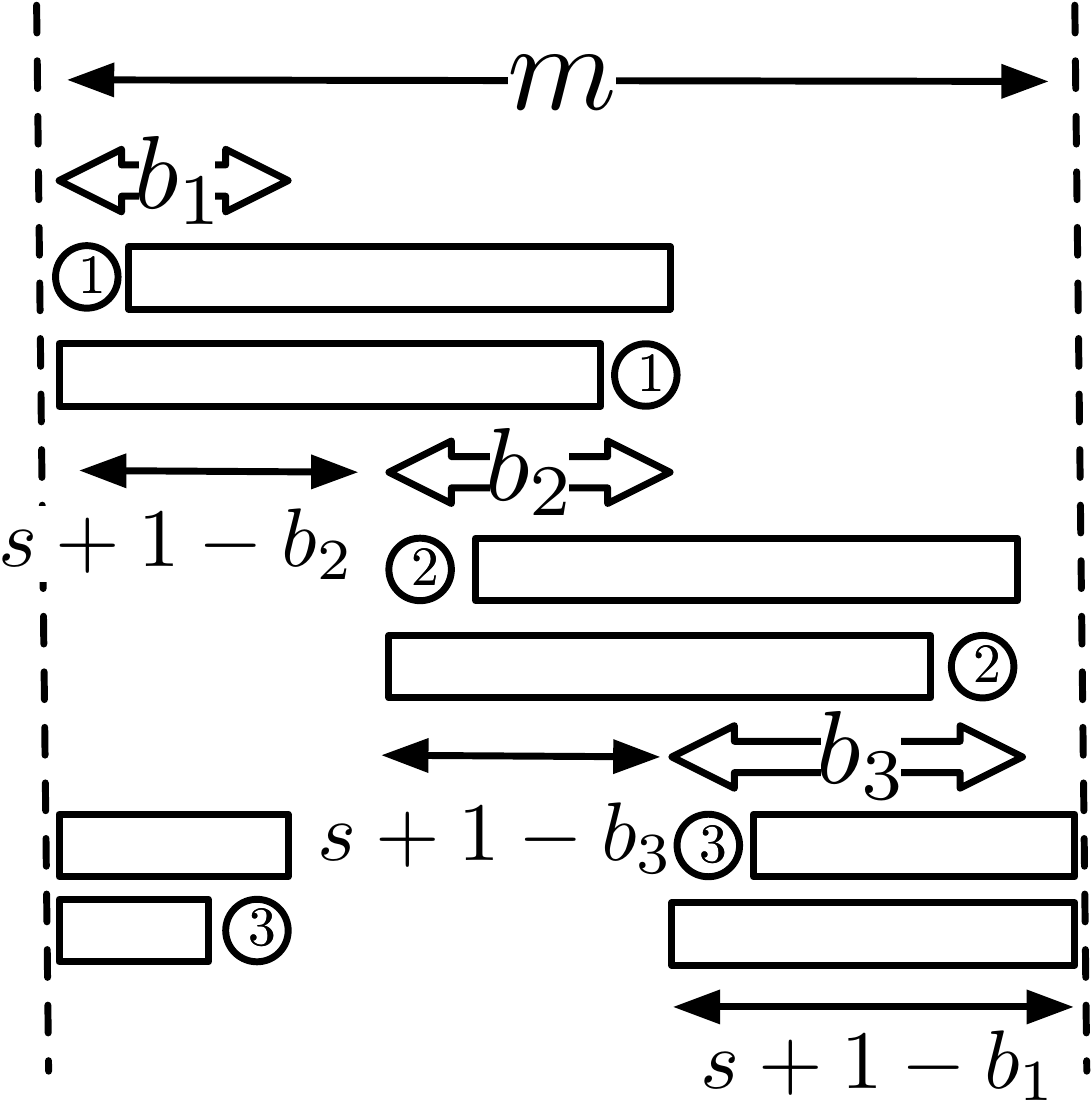}
  \caption{Codewords and corresponding desired message pairs for the converse bound.}
  \label{fig:desired_message_pairs}
\end{figure}

Fig.~\ref{fig:desired_message_pairs} illustrates the users, their desired messages, and sent codewords:
the users are denoted by their side information set that is represented as a rectangle;
the desired messages are devoted by the circled numbers next to a user; and
the codewords are denoted by arrows, 
with their ranges shown over the arrows.
Without loss of generality, assume $a_1=1$.
By the first transmission, $u_{b_1-s+1 \mod{m}}$ determines the desired message by $u_i$ to be $w_{b_1}$, and the desired message by $u_2$ to be $w_1$.
Note that the desired messages of these two users are adjacent to their side information sets. 
Therefore, these two users can mimic other users by decoding their desired messages. 
Specifically, $u_{b_1-s+1 \mod{m}}$ can mimic $u_{b_1-s+2 \mod{m}}$ and $u_2$ can mimic $u_1$.
To satisfy the security constraint, the desired messages of $u_{b_1-s+2 \mod{m}}$ and $u_1$ 
must be $d_{b_1-s+2 \mod{m}} = {b_1-s+1 \mod{m}}$ and $d_1=s+1$, respectively.

Now consider $u_1$. 
To satisfy $u_1$, there must exist a vector $v_1\in \mathsf{Span}(G)$ such that its $(s+1)$-th element is nonzero and all its nonzero elements are in the range $[1:s+1]$.
By a linear combination of $v_1$ and $g_1$ (the first row of the generator matrix), we can generate a vector $g_2$ such that its $1$st element is zero, its $(s+1)$-th element is nonzero,  and all its nonzero elements are in the range $[2:s+1]$.
Therefore, $g_2$ and $g_1$ are linearly independent, and  $g_2$ is a valid row of $G$.
The transmission of $g_2$, besides satisfying user $u_1$, also satisfies user $u_{s-b_2+3}$, whose decoded message must be $w_{s-b_2+2}$.
After decoding its desired message, user $u_{s-b_2+3}$ can mimic $u_{s-b_2+2}$ thus the desired message of $u_{s-b_2+2}$ must be $w_{2s-b_2+1}$.
Now we focus on $u_{s-b_2+2}$. 
By a similar argument, we can see there must exist a $g_3$ such that its nonzero elements are in the range $[s-b_2+3:2s-b_2+1]$ and the $(2s-b_2+1)$-th element is nonzero. 
Vector $g_3$ leads to the determination of the desired messages of another two users in the system and then of vector $g_4$ for another transmission.

We can see that the desired messages we found by this argument can be grouped into pairs. 
For instances, $(w_1, w_{s+1})$ and $(w_{s-b_2+2}, w_{2s-b_2+1})$ are two such pairs, denoted as the circled `1' and `2'  in Fig.~\ref{fig:desired_message_pairs}.
The two messages in a pair are 
at a distance of $s+1$.
The $i$-th and $(i-1)$-th pairs have a range overlap of size $b_i$.
The desired message pairs and the codewords form a `chain': the desired message pair requires a codeword; the codeword then determines the new desired message pair based on its range. 
In this way, the `chain' keeps until the newly determined desired message has already been generated.
For instance, in Fig.~\ref{fig:desired_message_pairs}, the message $w_{b_1}$ is a desired message generated by the first transmission;
since it is generated again by the third transmission, the chain stops with three transmissions.
The adjacent desired message pairs have a shift of $s+1-b_i$, shown in Fig.~\ref{fig:desired_message_pairs} as the solid arrows. 
When the chain stops, all the pairs' shifts sum up to at least $m$.
Let $k$ be the number of transmissions we discover in the argument. 
We have  $\sum_{i=1}^k (s+1-b_i) \geq m$.
Since $\ell \geq k$, we can replace $k$ with $\ell$ and have the inequality
\begin{align}
    s \ell +\ell \geq \sum_{i=1}^\ell b_i +m. 
    \label{eq:chain_count_ineq}
\end{align}

\paragraph{Maximum Number of Satisfied Users}
Here we consider the maximum number of users that can be satisfied by a given $G$.
For one transmission of range $b$, 
the maximum number of users that can be satisfied is $2(b-1)$. 
Let the set of users whose side information sets intersect but not contain the range of $g_i$ be denoted by $Q_i$, $|Q_i|\leq 2(b_i-1)$. 
We argue that all satisfied users must be in one of the $Q_i,i\in[\ell]$, i.e., $|\cup_{i=1}^\ell Q_i|=m$. 
This can be proved by contradiction. 
Without loss of generality, let $u_1$ be a user that is securely satisfied by $G$ but not in any $Q_i$.
Thus $G$ is a block diagonal matrix, and
$\mathsf{Span}(G)$ can be divided into two orthogonal spaces $\mathsf{Span}(G_1)$ and $\mathsf{Span}(G_2)$.
$\mathsf{Span}(G_1)$ is not helpful for $u_1$. 
In order for $u_1$ to decode, there must be a standard basis vector in $\mathsf{Span}(G_2)$ and thus there is a standard basis contained in $\mathsf{Span}(G)$.
This violates the security constraint in~\cite[Proposition~1]{individual_sec_picod}. 

Therefore, by union bound we have $m = |\cup_{i=1}^\ell Q_i| \leq \sum_{i=1}^{\ell} |Q_i| \leq 2\sum_{i=1}^\ell (b_i-1)$.
Finally, we have 
\begin{align}
    \sum_{i=1}^\ell b_i \geq \frac{m}{2}+\ell. \label{eq:user_count_ineq}
\end{align}

\paragraph{Final Step}
By combining~\eqref{eq:chain_count_ineq} and~ \eqref{eq:user_count_ineq} we have
\begin{align}
    \ell\geq \frac{3m}{2s},
\end{align}
which is the desired converse bound in Theorem~\ref{thm:gap_3}.

\section{Achievable schemes}
\label{sec:achievability}
When $\frac{m}{m-s}\in\mathbb{Z}$ we use the achievable scheme for the decentralized PICOD in~\cite{decentralized_picod}. 
The scheme satisfies the security constraint since all users have all but one messages that are involved in the codewords. Therefore, it is impossible for a user to decode more than one message.
The scheme is optimal since it is optimal without security constraint. 

Next we provide achievable schemes for two cases where our converse under linear encoding constraint is tight. 
One of the schemes can be further generalized to all even $m$. 
Lastly, we provide a class of feasible cases with odd $m$ and  $\frac{m}{m-s}\notin \mathbb{Z}$.

\paragraph{$\frac{m}{2s}\in \mathbb{Z}$}
\label{par:tight}
An optimal scheme is to send $\{w_{1+2sk}+w_{2+2sk}, w_{3+2sk}+w_{s-2+2sk}, w_{s-3+2sk}+w_{s-4+2sk}\}, k\in\{0,1,\dots,\frac{m}{2s}-1\}$, for a total of $\frac{3m}{2s}$  transmissions. 

\paragraph{$s=3$ and even $m$}
\label{par:s=3}
An optimal achievable scheme is $\{w_{1+2k}+w_{2+2k}\}, k\in \{0,1,\dots, \frac{m}{2}\}$, for a total of $m/2$ transmissions.
Note that in this case $\frac{3m}{2s}=\frac{m}{2}$.

\paragraph{$m$ is even}
\label{par:even_m}
We can further generalize the scheme for $s=3$, even $m$  case to all even $m$ cases.
The scheme transmits $\{\sum_{i=1+2k}^{s-1+2k \mod{m}} w_i\}, k\in \{0,1,\dots, \frac{m}{2}+1-\lceil \frac{s}{2} \rceil\}$.

\paragraph{odd $m$, $s=m-3\geq 8$, $\frac{m}{m-s}\notin \mathbb{Z}$}
This case shows that not all odd $m$ and $\frac{m}{m-s}\notin \mathbb{Z}$ are infeasible. 
The scheme has four transmissions, $\{ w_1+\sum_{i=1}^{\frac{m-7}{2}}w_{2i}+w_{m-6}, \sum_{i=1}^{\frac{m-9}{2}} w_{2i+1}+w_{m-5}+w_{m_4}, w_{m-3}+w_{m-2}, w_{m-1}+w_{m} \}$.
which are currently working to generalize.

\bibliographystyle{IEEEtranS}
\bibliography{refs}

\end{document}